\documentclass[twocolumn,aps,prb]{revtex4}

\usepackage{graphicx}

\begin{document}


\title{Raman study of carrier-overdoping effects on the gap in high-T$_c$ superconducting cuprates}
\author{T. Masui} 
\altaffiliation{JSPS Research Fellow}
\email{masui@istec.or.jp}
\author{M. Limonov}
 \altaffiliation{Permanent address: A. F. Ioffe Physical-Technical Institute, 
194021 St. Petersburg, Russia.}
\author{H. Uchiyama}
 \altaffiliation{JSPS Research Fellow}
\author{S. Lee}
\author{S. Tajima}

\affiliation{Superconductivity Research Laboratory, ISTEC, 1-10-13 Shinonome, Tokyo, 135-0062, Japan}
\author{A. Yamanaka}
\affiliation{Chitose Institute of Science and Technology, Chitose, Hokkaido 066-8655, Japan}
\date{\today}

\begin{abstract}
Raman scattering in the heavily overdoped (Y,Ca)Ba$_2$Cu$_3$O$_{7-\delta}$  
(T$_c$ = 65 K) and 
Bi$_2$Sr$_2$CaCu$_2$O$_{8+\delta}$ (T$_c$ = 55 K) crystals has been 
investigated. 
For the both crystals, the electronic pair-breaking peaks 
in the A$_{1g}$ and B$_{1g}$ polarizations were largely shifted 
to the low energies 
close to a half of 2$\Delta_0$, $\Delta_0$ being the maximum gap. 
It strongly suggests $s$-wave mixing into the $d$-wave 
superconducting order parameter and the consequent
 manifestation of the Coulomb screening effect 
in the B$_{1g}$-channel. 
Gradual mixing of $s$-wave component with overdoping is not due to
the change of crystal structure symmetry but a generic feature 
in all high-T$_c$ superconducting cuprates.

\end{abstract}


\maketitle

In spite of a tremendous amount of studies on the electronic phase diagram 
of high-T$_c$ superconducting cuprates (HTSC), 
there are many unresolved problems.
One of the remaining problems is the electronic state in the overdoped regime, 
where the transition temperature T$_c$ decreases with carrier doping. 
A naive picture is that the T$_c$ suppression is due to the weakening of 
pair-interaction such as electron correlation and thus the gap decreases 
in scale with T$_c$. 
However, the experimental results do not support such 
a simple picture.   
The recent STM study demonstrated a phase separated state 
in the overdoped regime, consisting of the superconducting and the normal 
metallic phase \cite{Davis}. 
This accounts for the anomalous increase in unpaired carriers \cite{Schutzmann}
and /or the decrease in superfluid density \cite{Uemura,Niedermayer} 
with overdoping. 
In this phase separation model, 
the nature of superconducting gap would be maintained. 
On the other hand, in the model assuming the quantum critical point 
at hole concentration $p \sim$0.19 \cite{Tallon}, 
it is expected that a gap nature radically changes 
in the heavily overdoped state where the pseudogap disappears. 
Therefore, the study of superconducting gap properties 
in the overdoped regime, 
such as a gap magnitude and symmetry, 
is of great importance to discuss the model for the electronic state 
as well as the pairing mechanism of HTSC. 

Most of the previous gap-studies in the overdoped regime are 
on Bi$_2$Sr$_2$CaCu$_2$O$_{8+\delta}$ (Bi-2212) by tunneling and 
Raman scattering spectroscopies \cite{Renner,Hackl2,Kendziora,Sugai}. 
Although it is almost established that the gap magnitude decreases 
with doping, the values are still controversial. 
For example, the pair-breaking peak energy (26 meV) observed 
in the Raman scattering spectrum of overdoped Bi-2212 
with T$_c$ = 57 K \cite{Kendziora} is much smaller than the value 
(2$\Delta_0$= 42 meV) obtained from the tunneling 
experiment for nearly the same T$_c$ sample \cite{Renner} .  
It has also been suggested by the Raman study 
that the gap-T$_c$ ratio, 2$\Delta/k_B T_c$, 
radically decreases with overdoping.  
For example, in Tl$_2$Ba$_2$CuO$_{6+\delta}$ (Tl-2201), 
it changes from 2$\Delta/k_B T_c$=8 at the optimum doping down to less 
than 3 at the heavy overdoping with $T_c/T_{c,max}$ = 0.4 \cite{Gasparov}.

Another problem is on the gap symmetry,
which is crucially important to discuss the pairing mechanism. 
Raman scattering has been playing an important role 
in establishing the $d$-wave symmetry of HTSC \cite{Hackl}. 
However, in some of the overdoped HTSC, 
the pair-breaking peak 
in Raman spectra seems to change its polarization dependence
that is characteristic to $d$-wave superconductor \cite{Hackl2}. 
Although in most cases it has been attributed to the $s$-wave component 
mixing due to orthorombicity \cite{Maki,Annett,Willemin}, 
it is necessary to examine whether it can be a more generic nature in HTSC 
or not.

Motivated by these problems, we have investigated the Raman scattering spectra 
of overdoped HTSC. 
In this work, we present Raman scattering spectra of heavily overdoped 
(Y,Ca)Ba$_2$Cu$_3$O$_{7-\delta}$ (Y/Ca-123) and 
Bi$_2$Sr$_2$CaCu$_2$O$_{8+\delta}$ (Bi-2212), focusing on the pair 
breaking peaks. 
It was found that for these heavily overdoped crystals the electronic 
pair-breaking peaks for all polarizations are located at much lower energy 
than 2$\Delta_0$. 
These Raman data are well explained by the $s$-wave mixing with overdoping, 
accompanied by the change of the screening effect, 
as proposed by Nemetschek {\it et al.} \cite{Nemetschek}. 

The Y/Ca-123 crystals were grown by a crystal pulling technique \cite{Yamada} 
and detwinned under uniaxial pressure. 
By annealing further in high-pressure (800 atm) oxygen at 450 $^\circ$C, 
we prepared heavily overdoped crystals with T$_c$=65 K. 
The Ca-content (12 \%) and the oxygen content 7-$\delta$=6.87 were determined 
by an inductively coupled plasma analysis and iodometric titration, 
respectively. 
The Ca-free Y-123 crystal with T$_c$ = 93 K was also prepared for a reference 
of optimally doped sample. 
The Bi-2212 crystals were grown 
by a KCl flux technique. 
The T$_c$ for the as-grown crystals was about 90 K,  
close to the values reported for optimally doped Bi-2212.
Heavily overdoped crystals with T$_c$ = 55 K  were prepared 
by post-annealing of the as-grown crystals at 450 $^\circ$C for 100 hrs. 
under oxygen gas pressure of 900 atm. 

The Raman scattering spectra were measured in the pseudo-backscattering 
configuration with a T64000 Jobin-Yvon triple spectrometer equipped 
with a liquid-nitrogen cooled CCD detector. 
A typical spectral resolution was 3 cm$^{-1}$. 
The Ar$^+$-Kr$^+$ laser line of 2.54 eV was used for Y/Ca-123,
while several excitation energy E$_{exc}$ ranging from 1.9 eV to 2.7 eV were used for Bi-2212. 
The power was about 10 mW, focused to a spot of about 0.4 $\times$ 0.5 mm$^2$
on the sample surface. 
The overheating was estimated to be less than 10 K. 
All the measured spectra were corrected for the spectrometer sensitivity 
by comparison with that of BaF$_2$ and the contribution of the Bose factor 
has been removed. 
For low-temperature measurements, a closed-cycle cryostat 
was used with temperature stabilization better than 1 K.

Although the crystal structures of Y/Ca-123 and Bi-2212 are orthorhombic, 
hereafter all symmetries refer to a tetragonal D$_{4h}$ point group, 
as the tetragonal treatment for the CuO$_2$-plane has been commonly adopted
and proved to work well in the previous studies \cite{Hackl}. 
For Y/Ca-123 crystals,  we extracted 
B$_{1g}$(X\ensuremath{'}Y\ensuremath{'}), 
B$_{2g}$(XY), and A$_{1g}$(X\ensuremath{'}X\ensuremath{'}-XY) Raman responses. 
For Bi-based superconductors, X and Y are indexed along the Bi-O 
bonds, rotated by 45\ensuremath{^\circ} relative to the Cu-O bonds. 
Therefore, for Bi-2212 crystals the above three Raman 
channels were taken as B$_{1g}$(XY), B$_{2g}$(X\ensuremath{'}Y\ensuremath{'}), 
and A$_{1g}$(XX-X\ensuremath{'}Y\ensuremath{'}).

Figure 1 shows the B$_{1g}$-, A$_{1g}$-, 
and B$_{2g}$- Raman spectra of Y/Ca-123 
single crystals above and below T$_{c}$.
The phonon peaks' profile is not much different from that of Y-123, 
except for a decrease of peak intensity in B$_{1g}$-phonon 
at $\sim$340 cm$^{-1}$. 
In the electronic continuum, 
a broad bump appears at around 220 cm$^{-1}$ below T$_c$,
which can be ascribed to a pair-breaking peak.
Note that the pair-breaking peak intensity is weaker 
in the A$_{1g}$-spectrum than in the B$_{1g}$. 
The electronic Raman responses were extracted by fitting in the same way 
as in ref. \cite{Limonov2000},
where electron-phonon coupling is taken into account.
The obtained spectra are plotted in Fig.2 together with the spectra of 
optimally doped pure Y-123 (T$_c$ = 93 K) in comparison. 
The spectra of Y-123 are in good agreement 
with the previously reported results \cite{Cardona}. 

In considering the overdoping effect, there are three remarkable features 
in Figs. 1 and 2. 
First, both for the A$_{1g}$- and B$_{1g}$-polarizations 
the pair-breaking peak  energies (E$_p \approx$ 220 cm$^{-1}$) 
are substantially 
lower in overdoped Y/Ca-123 than those in optimally doped Y-123. 
A weaker but similar tendency was reported for the B$_{1g}$-spectra of 
slightly overdoped 
(Y$_{0.95}$Ca$_{0.05}$)Ba$_{2}$Cu$_{3}$O$_{7-}$$_{\ensuremath{\delta}}$ with 
T$_{c}$=82.7 K \cite{Bock} where E$_p$(B$_{1g}$) $\approx$ 300 cm$^{-1}$. 
Second, the symmetry dependence of the pair-breaking peak energy E$_p$ 
is significantly weakened
in the overdoped Y/Ca-123, 
the B$_{1g}$ and A$_{1g}$ pair-breaking peaks being located close to each other
at around 220 cm$^{-1}$. 
As to the B$_{2g}$ spectrum, 
the intensity is too weak to conclude the pair-breaking peak position.
Third, the height of A$_{1g}$ pair-breaking peak is suppressed 
in overdoped Y/Ca-123, while a large bump is observed 
in optimally-doped Y-123 \cite{peak-height}. 

Similar overdoping effects are observed in Bi-2212.
Figure 3 illustrates the Raman spectra in A$_{1g}$, B$_{1g}$, and B$_{2g}$ 
channels for the overdoped Bi-2212 crystals. 
To suppress the phonon features in the B$_{1g}$ and B$_{2g}$, 
the laser light with E$_{exc}$=1.92 eV was used \cite{Yamanaka}. 
The pair-breaking peaks are observed at the polarization independent 
energy of about 180 cm$^{-1}$,
which is in sharp contrast to the polarization dependent peak energies 
in optimally doped Bi-2212 crystals \cite{Staufer}. 
It is clearly demonstrated in Fig.4, where pure 
superconducting response I(10K)-I(T$_c$) is presented. 
In the case of optimally doped Bi-2212 with T$_c$=90K, 
the peaks were observed at about 520 cm$^{-1}$ in the B$_{1g}$ 
and 390 cm$^{-1}$ in the A$_{1g}$ spectra. 
Note that E$_p$(B$_{1g}$) in overdoped Bi-2212
becomes one third of the value for optimally doped samples,
although the T$_c$ is suppressed by only 30\%. 
The weakened polarization dependence of E$_p$ and the radical decrease 
in E$_p$ were reported also for the famous overdoped material 
Tl-2201 \cite{Kendziora,Gasparov},
and thus we believe that this is a common property in all overdoped HTSC.

For the loss of polarization dependence of E$_p$,
three possible origins can be considered.
First, the gap symmetry changes to $s$-wave,
and E$_p$ indicates 2$\Delta_s$,
where $\Delta_s$ is an $s$-wave gap energy.
Second, the screening effect disappears in the A$_{1g}$-spectrum,
as is observed in three layer compounds like 
Bi$_{2}$Sr$_{2}$Ca$_{2}$Cu$_{3}$O$_{10+}$$_{\ensuremath{\delta}}$ (Bi-2223) 
\cite{Limonov2002}.
Here E$_p$ corresponds to 2$\Delta_0$,
$\Delta_0$ being the maximum amplitude of $d$-wave gap.
Third, the $s$-wave component is mixed into the $d$-wave,
introducing the screening effect in the B$_{1g}$-spectrum.
In this case, E$_p$ becomes smaller than 2$\Delta_0$,
as demonstrated by Nemetschek \textit{et al}.\cite{Nemetschek}

The first possibility of $s$-wave gap was proposed 
by Kendziora {\it et al.} \cite{Kendziora}.
However, it is unlikely because 
the low-$\omega$ Raman scattering shows the polarization dependence
typical for a $d$-wave superconductor.
As seen in Fig.3 and in previous reports \cite{Hewitt}, 
linear \ensuremath{\omega} 
- law for the A$_{1g}$ -- and B$_{2g}$ -- polarizations and  \ensuremath{\omega}$^{3}$-like-law for the B$_{1g}$ polarization were observed. 
These are in general agreement with the theoretical predictions for 
\textit{d}-wave pairing symmetry \cite{Hackl,Gatt} 
but in contrast to the results \cite{Kendziora} 
reporting polarization independent low-frequency profiles. 

In order to distinguish the remaining two cases, 
it is crucial to know the maximum gap energies $\Delta_0$. 
For the overdoped Bi-2212,
the gap values are available from ARPES \cite{Feng} 
and scanning tunneling microscope (STM) studies \cite{Renner}.
According to the STM result on the Bi-2212 crystal with T$_c$ =56 K that 
is similar to ours,
the gap energy $\Delta_0$ is 21 meV ($\approx$ 170 cm$^{-1}$) \cite{Renner}.
The ARPES on the Bi-2212 crystal with T$_c$ = 65 K also suggests that
$\Delta_0$ = 20 meV ($\approx$ 160 cm$^{-1}$) \cite{Feng}.
It turns out that the B$_{1g}$-peak energy ($\sim$ 180 cm$^{-1}$) 
observed in the present 
Raman study is closer to $\Delta_0$ rather than 2$\Delta_0$.
This fact can exclude not only the first possibility of $s$-wave gap
but also the second possibility that 
the screening effect disappears in the A$_{1g}$-channel.
In the latter case, 
the A$_{1g}$- and B$_{1g}$-peaks should be located at 2$\Delta_0$.
Another support to exclude the second possibility is the weak intensity 
of the A$_{1g}$ pair-breaking peak which should be strong 
in the case of screening-free case. 

The third possibility was proposed by Nemetschek \textit{et al}. 
\cite{Nemetschek}.  
For fully oxygenated Y-123, $s$-wave mixing was indicated by the XY-anisotropy 
of gap amplitude in Raman \cite{Limonov2000} and ARPES \cite{Lu} studies 
and by the gap peak splitting in tunneling spectra \cite{Yeh}. 
For a cylindrical Fermi surface, when the $s$-component is mixed, 
the order parameter is given by \cite{Nemetschek,Maki}: 
$\Delta({\bf k}) = \Delta_d[cos(2\phi) + r]$. 
Here $\phi$ is the angle that $\textbf{k}$ makes with the \textit{a} 
axis in the \textit{ab} plane and \textit{r} represents the $s$-component 
mixing rate \cite{comment}. 
The gap amplitude $|\Delta({\bf k})|$ shows double maxima $\Delta(1\pm |r|)$.
Correspondingly, unscreened Raman response has two peaks at 2$\Delta_0$ 
= $2\Delta_d(1+|r|)$ and 2$\Delta_d(1-|r|)$. 
In the case of pure $d$-wave with $|r|$=0, 
the screening effect manifests itself only in A$_{1g}$ channel, 
while for the $d$+$s$ case, 
admixture of an $s$ component to the predominant $d$ 
component introduces the screening effect also in the B$_{1g}$-channel. 
When $|r|$ becomes large,
the higher energy peak in B$_{1g}$-spectrum is smeared out by the screening,
leaving the smaller energy peak as a predominant peak in B$_{1g}$.
In this case,
E$_p$ corresponds to 2$\Delta_d(1-|r|)$.
This feature appears to be unique to the $d$+$s$ wave superconductor, 
among possible candidates for order parameter in HTSC like $d$+$is$ and $d$+$g$ \cite{Nemetschek}. 
The present result for the overdoped Bi-2212 corresponds to the case 
for $|r|$ $\sim$ 0.3,
in which the peak energies of B$_{1g}$, A$_{1g}$ and B$_{2g}$-channels
are close to $\Delta_0$ \cite{Nemetschek}.
In fact, E$_p$(B$_{1g}$) $\approx$ E$_p$(A$_{1g}$) $\approx$ 180 cm$^{-1}$ 
is close to $\Delta_0$=170 cm$^{-1}$ determined by the STM measurement 
on the Bi-2212 crystal with nearly the same T$_c$ as ours \cite{Renner}. 
As for our overdoped Y/Ca-123 crystal, 
we have measured angle-resolved photoemission spectra (ARPES) \cite{Uchiyama}, 
and observed a gap about 18$\pm$2 meV ($\approx$ 145$\pm$15 cm$^{-1}$) 
near the X-point, while  
a newly developed 1-D state conceals the gap structure near the Y-point. 
On the analogy to the overdoped Y-123 \cite{Lu}, the gap near the 
X-point is expected to reflect the smaller gap $\Delta_d(1-|r|)$ 
as E$_p$(B$_{1g}$) does in Raman spectrum. 
Thus, $|r|$ cannot be estimated in this case. 
However, judging from the very close positions of the A$_{1g}$- and 
B$_{1g}$-pair-breaking peaks, we speculate that $|r|$ is larger 
than 0.15 \cite{Nemetschek} but close to 0.3. 

As an example of the screening free case, 
we show  the A$_{1g}$- and B$_{1g}$- superconducting 
response of optimally doped Bi-2223 
with T$_{c}$ = 109 K at the bottom of Fig.4 \cite{Limonov2002}. 
Here, the Coulomb screening in the A$_{1g}$-channel disappears
owing to the band splitting specific 
to a multiple layer superconductor \cite{Krantz},
giving E$_p$(B$_{1g}$) = E$_p$(A$_{1g}$) 
= 580 cm$^{-1}$ \ensuremath{\approx} 2\ensuremath{\Delta}$_{0}$. 
This interpretation of E$_p$(A$_{1g}$) = 2\ensuremath{\Delta}$_{0}$
was supported by the ARPES-data for the same compound \cite{Muller}, 
where the maximum gap energy \ensuremath{\Delta}$_{0}$ of about 37 meV (2\ensuremath{\Delta}$_{0}$ 
= 590 cm$^{-1}$) was obtained. 
It should be noted that a strong intensity of the 
A$_{1g}$-peak in Bi-2223 is also the evidence 
for the disappearance of screening effect,
which is in sharp contrast to the very weak A$_{1g}$-peak 
in the overdoped Bi-2212 and Y/Ca-123.
Disappearance of screening effect due to band splitting was first observed 
in Bi-2223 under the resonance condition at the orange-red excitation 
(E$_{exc}$ $<$ 2.3eV) \cite{Limonov2002}. 
A similar strong enhancement of A$_{1g}$ pair-breaking peak 
was also observed in Hg-1223 and some other 
multi-layer cuprates\cite{Limonov2002}.
On the other hand, although we examined several excitation energies for the measurement of Bi-2212 spectra,
neither the resonance phenomenon nor the band splitting effect was observed
within the present energy range.

The most intriguing problem is the origin of the \textit{s}-component in overdoped HTSC. 
Although the orthorhombic distortion in Y-123 
crystals favors the $d$+$s$ mixing \cite{Maki,Donovan}, 
this cannot explain the present Raman results. 
First, the orthorhombicity is smaller 
in our overdoped Y/Ca-123 crystals than 
in the fully oxygenated Y-123.
Nevertheless, the $s$-wave mixing rate $|r|$ is larger in the former 
than in the latter ($|r| \sim$ 0.15) \cite{Nemetschek}.
Therefore, orthorhombicity does not correlate with the $s$-wave mixing rate.

Second, the Raman spectra of the less orthorhombic Bi-2212 
and the tetragonal Tl-2201 also require the $d$+$s$ model, 
although the $d$+$s$ symmetry is less favorable for the 
tetragonal structures \cite{Annett}. 
Therefore, it is necessary to consider that
the $s$-wave mixing is of purely electronic origin linked to overdoping. 
If the inhomogeneous electronic state induces local lattice distortion, 
it might introduce $s$-wave component. 
It is interesting to examine a more radical scenario 
that the pairing mechanism itself changes with overdoping. 

In summary, we investigated Raman scattering spectra of two 
heavily overdoped compounds, Y/Ca-123 (T$_{c}$ = 65K) and Bi-2212 
(T$_{c}$ = 55K). 
Comparing the results of ARPES and STM with our Raman results,
we found that the pair-breaking peaks for B$_{1g}$- and A$_{1g}$-polarizations 
are located at the energies substantially lower than 2$\Delta_0$,
just as predicted theoretically for $d$+$s$-wave superconductor. 
The change of the gap symmetry seems to be intrinsic to the overdoped regime in HTSC.
The present findings, a drastic change of the gap nature 
in the overdoped regime of HTSC, 
must be a crucial test for the theoretical models 
for the electronic phase diagram of HTSC as well as the high-T$_c$ mechanism. 
Further theoretical works to explain the mechanism of $s$-wave mixing 
are desired.
The authors appreciate K. Kuroki at the Univ. Electro-Communications 
for useful discussions.
This work is supported by New Energy and Industrial Technology 
Development Organization (NEDO) as Collaborative Research and 
Development of Fundamental Technologies for Superconductivity 
Applications. 

\newpage
\begin{figure}
 \begin{center}
   \includegraphics[width=6cm]{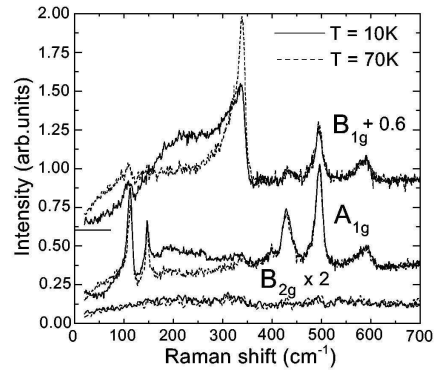}
  \end{center}
\caption{Raman spectra of overdoped Y/Ca-123 
crystal in B$_{1g}$-, A$_{1g}$-, and B$_{2g}$ - polarizations above and 
below T$_{c}$ = 65 K. B$_{1g}$- spectra are shifted in the vertical scale 
by 0.6.}
\end{figure}
\begin{figure}
 \begin{center}
   \includegraphics[width=6cm]{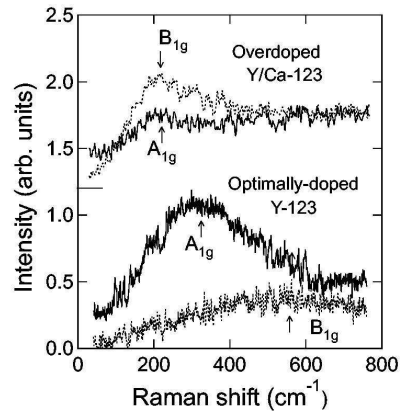}
  \end{center}
\caption{The electronic component of overdoped Y/Ca-123 and optimally 
doped Y-123 crystals in B$_{1g}$-, A$_{1g}$- polarizations at T=10 K. 
The spectra of Y/Ca-123 are shifted 
in the vertical scale by 1.2. }
\end{figure}
\begin{figure}
 \begin{center}
   \includegraphics[width=6cm]{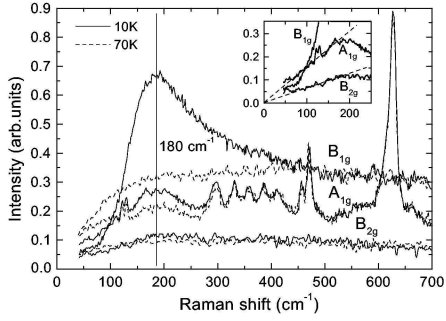}
  \end{center}
\caption{Raman spectra of heavily overdoped Bi-2212 crystals 
in B$_{1g}$-, A$_{1g}$-, and B$_{2g}$ - polarizations above and below T$_{c}$ 
= 55 K, measured at the excitation energy of 1.92 eV. 
Inset shows the low-$\omega$ parts at T=10 K.
}
\end{figure}
\begin{figure}
 \begin{center}
   \includegraphics[width=6cm]{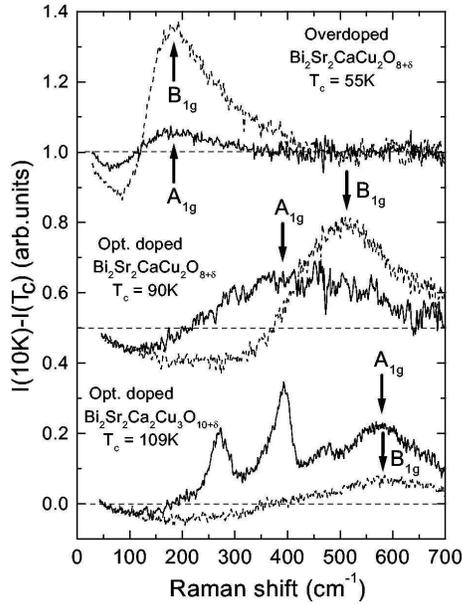}
  \end{center}
\caption{The A$_{1g}$- and B$_{1g}$- differential spectra I(10 K)-I(T$_{c}$) 
of overdoped Bi-2212, nearly optimally doped Bi-2212 and Bi-2223, 
measured at the excitation wavelength of 1.92 eV. 
In the A$_{1g}$-spectrum of Bi-2223, two phonons are apparent 
due to a strong resonance effect accompanied by the large 
phonon renormalization \cite{Limonov2002}. 
 }
\end{figure}
\end{document}